# White Rabbit Time Synchronization for Radiation Detector Readout Electronics

Wolfgang Hennig
XIA LLC
Hayward, CA, USA
whennig@xia.com

Shawn Hoover

shawn@shawnhoover.dev

***Abstract*— As radiation detector arrays in nuclear physics applications become larger and physically more separated, the time synchronization and trigger distribution between many channels of detector readout electronics become more challenging. Clocks and triggers are traditionally distributed through dedicated cabling, but newer methods such as the IEEE 1588 Precision Time Protocol and White Rabbit allow clock synchronization through the exchange of timing messages over Ethernet.**

**Consequently, we report here the use of White Rabbit in a new detector readout module, the Pixie-Net XL. The White Rabbit core, data capture from multiple digitizing channels, and subsequent pulse processing for pulse height and constant fraction timing are implemented in a Kintex 7 FPGA. The detector data records include White Rabbit time stamps and are transmitted to storage through the White Rabbit core's gigabit Ethernet data path or a slower diagnostic/control link using an embedded Zynq processor. The performance is characterized by time-of-flight style measurements and by time correlation of high energy background events from cosmic showers in detectors separated by longer distances. Software for the Zynq processor can implement "software triggering", for example to limit recording of data to events where a minimum number of channels from multiple modules detect radiation at the same time.**

***Keywords—White Rabbit, digital detector readout electronics, software triggering***

## I. Introduction

In[1] large nuclear physics applications, which can include hundreds or thousands of separate detector channels, it is usually required to synchronize the clocks of multiple digital detector readout electronics modules to ensure that data from different channels can be matched by their time stamps and that time differences can be computed to high precision, for example for time of flight measurements or coincidence windows. In addition, trigger signals may be distributed to collect data from "passive" channels based on interaction in "active" channels (e.g. a central contact triggering the segments of a segmented HPGe detector) or to only collect data from the detector array if a minimum number or a certain combination of channels have interactions. Such clocks and triggers are traditionally distributed through dedicated cabling from channels, which can become quite complex [1], [2], [3], [4], [5]. The timing requirements depend on the application, and can range from hundreds of nanoseconds for coincidence background measurements [6] to ideally tens of picoseconds for time-of-flight measurements. In practice noise, jitter and light collection in the detector or photomultiplier tubes (PMT) limit the performance to hundreds of picoseconds [7],[8],[9], while the electronics themselves can reach time resolutions below 10 ps with idealized signals from a pulser, even when digitizing at less than 1 GSPS [9].

More recently, time synchronization through data networks has been developed, such as the IEEE 1588 precision time protocol [10] (PTP) and in particular its "high accuracy profile" also known as White Rabbit [11]. Since detector readout electronics is usually connected to a data network in any case for readout to storage, these methods can be used as an alternative [6]. The challenge for the use of such network synchronization techniques in detector readout electronics is to integrate the synchronization with the data capture from analog to digital converters (ADC) and the processing of digitized detector signals in a field programmable gate array (FPGA). We report here how this has been implemented for a new electronics module, the Pixie-Net XL. We measure the performance of the electronics in a variety of jitter and resolution tests, and specifically characterize the timing performance in two types of measurements: 1) time-of-flight style measurements equivalent to two detectors exposed to coincident gamma radiation, and 2) correlating high energy background radiation from cosmic showers in two detectors separated by larger distances.

## II. System Description

### A. Hardware Design

The electronics hardware design centers on two Kintex 7 FPGAs. A block diagram of the design is shown in Fig.1. Each FPGA is connected with high speed LVDS lines, gigabit transceiver lines, and slower CMOS control lines to a high density connector for an ADC daughter card that implements multiple channels of analog signal conditioning and digitization. Implementing this circuitry on a daughter card facilitates customization of the inputs and digitization rate/precision for different applications. Three daughter cards have been designed so far: DB01 with a 75-125 MSPS, 14bit quad ADC for 4 channels with single ended coaxial input signals; DB02 with four 250 MSPS, 12-14bit dual ADCs for 8 channels with differential inputs bundled in a micro-HDMI connector; and DB06 with four 250 MSPS, 16bit ADCs for 4 channels with single ended coaxial input signals. Each FPGA is further connected to a dedicated 4 Gbit SDRAM memory for buffering of output data, an SFP+ card cage for Ethernet I/O, and a variety of general purpose I/O connections.

[1] This work was supported in part by the U.S. Department of Energy under Grant No. DE-SC0017223.

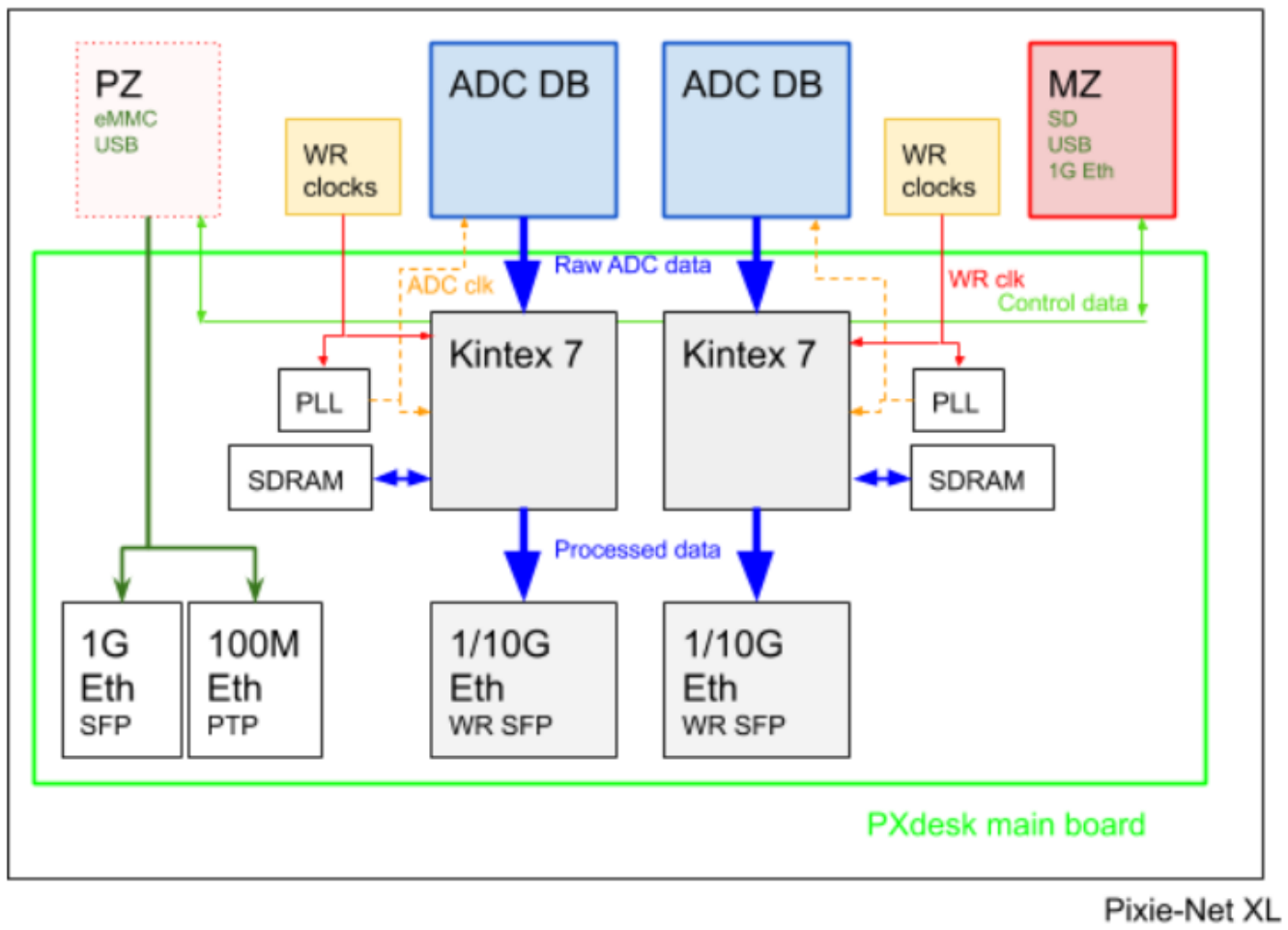

Fig. 1. Block diagram of the Pixie-Net XL. Note the high speed data flow is directly from ADC to FPGA to SFP (Ethernet) and is designed for up to 10 Gbit/s.

Each FPGA is also connected to two voltage controlled oscillators for clock adjustments based on the White Rabbit synchronization, generating a 125 MHz "main" clock for the 1G Ethernet interface and the White Rabbit core clock, and a 20 MHz or 62.5 MHz White Rabbit "helper" clock. This circuitry is implemented again on a separate daughtercard ("WR clocks" in Fig. 1), to easily implement changes for improved jitter or for generating of the different 10G base frequency (156.25 MHz). The "main" oscillators' output is copied and multiplied to the appropriate ADC sampling rate in dedicated phase lock loop circuitry (PLL). Both FPGAs are also connected to a slower control and setup bus from a Zynq processor module with its own Ethernet, USB, and memory peripherals (either Avnet MicroZed or PicoZed, "MZ" or "PZ" in Fig. 1). The Zynq processor is used to configure the FPGA, set parameters, and read out data in diagnostic mode. The Pixie-Net XL is designed as a desktop module with 12V DC power input; on-board regulators generate the various supply voltages required by the digital and analog circuitry.

### B. *Firmware Design*

The firmware of each FPGA is divided into 4 major sections: detector pulse processing, controller I/O, output data packaging, and the White Rabbit core.

The detector pulse processing consists of multiple identical logic modules to process the data streams of the different ADC channels, including sub modules for issuing triggers on the rising edge of a pulse, for capturing sums of a trapezoidal filter for pulse height measurements, for correcting the exponential decay and reconstructing pulse heights from the filter sums, for capturing full speed ADC waveforms, for capturing 3 short sums of programmable length and position before and after the rising edge, for computing constant fraction timing of the pulse arrival time, for latching internal and external time stamps, and for accumulating run statistics such as count times and input and output rates. The 3 short sums can be used for many types of pulse shape analysis, such as neutron/gamma discrimination for liquid scintillators. The functionality of the pulse processing is similar to that of other Pixie models [12], [13].

The controller I/O logic is implemented as a slow data bus between the FPGA fabric section of the Zynq processor and both Kintex 7 FPGAs. In each Kintex FPGA, it defines registers for pulse processing parameters, and makes the data captured from the detector signal available for readout in diagnostic mode. On the Zynq processor, the bus is connected to a xillybus lite core [14] to make FPGA fabric registers visible to the Zynq's ARM processor running Linux. Thus user software on the Linux side, written as simple C programs with file-like access to Zynq FPGA registers and indirect access to the registers of the Kintex FPGAs, can control the DAQ parameters and operation. In a similar manner, the Zynq processor controls slower I/O for the ADC daughter cards to set gains and offsets. Parameters are stored in settings files on the Zynq's SD card memory. The Linux functions further include a webserver and associated cgi programs to make output data available over the network for download or display the data in tables and graphs.

The data output section receives the captured event data from the pulse processing. Data from all processing channels is funneled into the SDRAM memory set up as a FIFO [15] and then assembled with IPv4 and UDP headers into UDP data packages for the Ethernet output, one captured event per package. The data flow is either directed by the Zync processor (for example advancing only those events that match a user defined coincidence pattern) or free flowing and entirely contained in the FPGA. A separate 4-bit data path transmits the measured pulse height values from FPGA to Zynq processor for the Zynq processor to increment an MCA energy histogram in its memory and periodically save it to file on the SD card memory.

The White Rabbit core is implemented as a direct copy of the reference design [16], with small adjustments for actual pinout of the chosen FPGA package. The synchronized White Rabbit date and time, 68 bits of seconds and nanoseconds, is available in the output registers visible to the Zynq processor, and can also be latched as an additional time stamp in the pulse processing logic for each event. The White Rabbit core controls the voltage controlled oscillators via two digital to analog converters, and the output of these oscillators is used for FPGA and ADC clocking as described above. The White Rabbit core also provides a data interface for "user data" to/from the network, receiving the UDP event packages described above. In alternative firmware, the White Rabbit core can be replaced with a generic 10G Ethernet core for higher data output rates (but without network time synchronization).

### C. *Software Design*

While basic operation of the Pixie-Net XL can be performed with the simple C programs running on the Zynq processor, we also developed in this project a concept of "software triggering", in which decisions to record specific events are not made via hardwired connections to the FPGAs, but instead through the exchange of data packets over the network by software. This can significantly reduce the cabling complexity for larger detector systems. In this approach, each Pixie-Net XL in a multi-module system sends out minimal data packages (metadata) containing timestamps, channel ID, and other essential information. This metadata is used by a central Decision Maker (DM) to make accept/reject decisions, which are then communicated back to all Pixie-Net XL modules. The Pixie-Net XL then independently move their full data to long term storage or discard.

For the actual implementation, we adopted a publish/subscribe, small packet message-oriented communication style based on the high performance open source messaging library ZeroMQ. ZeroMQ's PUB and SUB socket types, built on a foundation of automatic connection management, allow us to program in a decoupled publish/subscribe style while still leaning on the inherent properties of TCP transport for flow control and reliable ordered messaged delivery. We first tested the ZeroMQ library and benchmarked binaries on the earlier Pixie-Net electronics [6] and a variety of virtual and bare metal machines. Using the Pixie-Net as a server and a Ubuntu x86_64 virtual machine as a client exchanging 10,000 10-byte messages over a non-dedicated LAN, we observed one-way latencies ranging from 161-211 microseconds. (Lower latencies of 65-80 microseconds were observed on x86_64 machines alone.)

With the messaging library validated on the target platform, we developed the socket types and message values by prototyping DAQ and DM applications in Ruby with the ffi-rzmq library. The DM broadcasts run start synchronization commands and trigger decisions using a PUB socket and listens for trigger metadata using a PULL socket. The corresponding sockets in the DAQ are SUB and PUSH, respectively.

We then integrated the DAQ sockets into the Pixie-Net XL Zynq acquisition program and synchronized communication between two Pixie-Net XL and a Ruby prototype DM on a shared LAN. In the Zynq acquisition program, each Pixie-Net XL reads the metadata from the FPGA (while the full data is buffered in the SDRAM), sends it as a "trigger" to the DM, and also stores it locally in a ring buffer. The prototype DM receives the triggers and broadcasts accept responses for corresponding time ranges to both Pixie-Net XL. The Pixie-Net XL periodically poll for accept responses and match it to local data. If the time stamps of locally stored data fall in the acceptance time range, the corresponding full data packet in the FPGA's SDRAM FIFO is advanced to the White Rabbit Ethernet interface. Data from before the acceptance time range is discarded. Tracking trigger statistics in each node, we demonstrated end-to-end communication of triggers between multiple DAQs and the prototype DM with no processing delays or lost data at a modest input load.

## III. Characterization Measurements

### A. *Software Reported Timing Precision*

The White Rabbit "SoftPLL" core reports a number of performance values, in particular the clock offset of WR slave to WR master. The clock offset is reported to a GUI screen or to a log file about once per second. Log files were recorded over a period of time, typically 500-1000s, and the clock offset values were histogrammed into timing distributions, for which a FWHM timing resolution can be computed with a Gaussian fit. Such tests were performed with either a Pixie-Net XL, a commercial WR switch (Seven Solutions WRS 3/18), or a commercial WR-LEN daisy-chain module (Seven Solutions) as the clock master and either a Pixie-Net XL or a WR-LEN as the clock slave. Timing resolutions are in the range of 4.3 – 15.3ps FWHM, as shown in Fig 2.

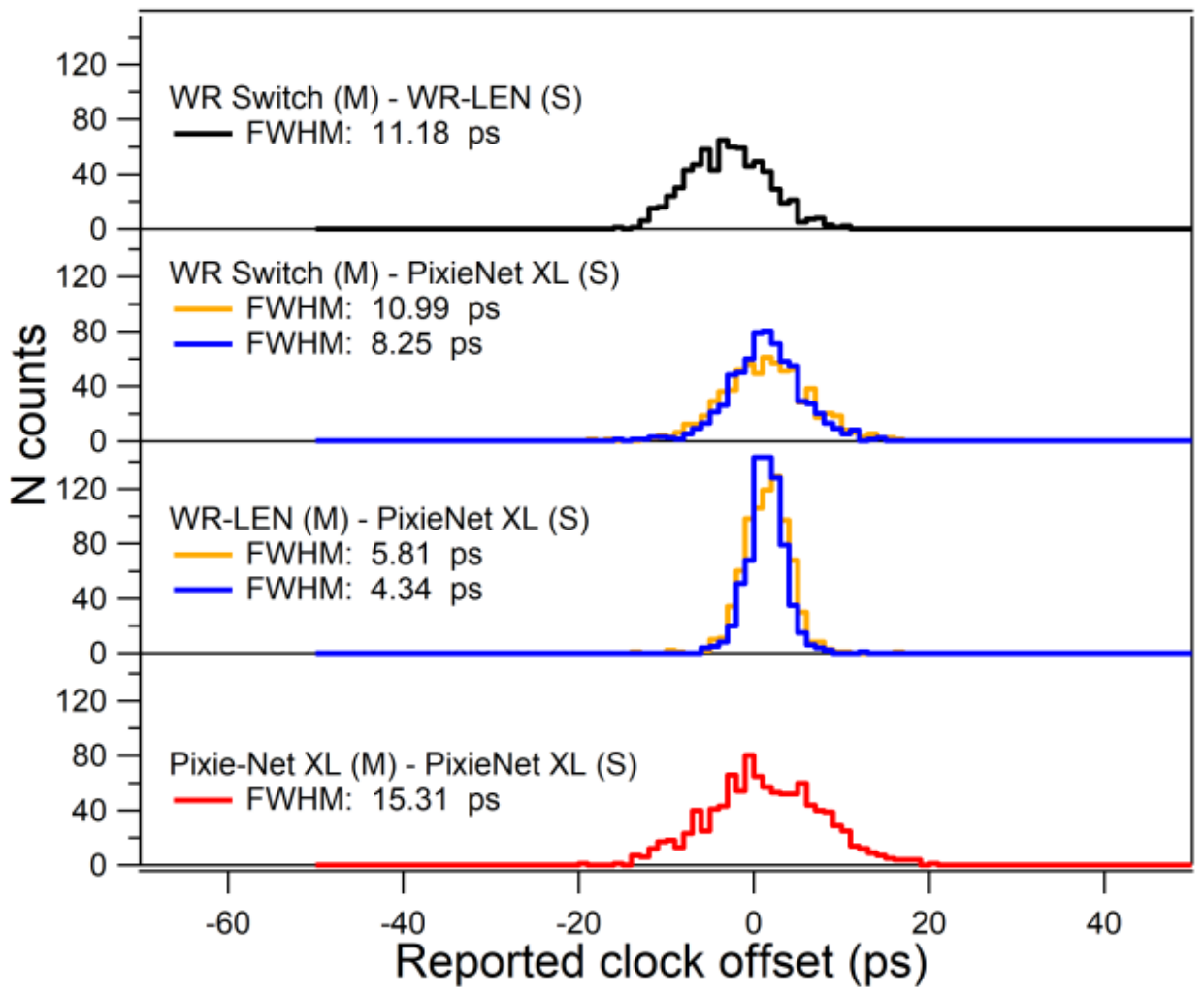

Fig. 2. Histogrammed reported offset from master clock.

### B. *Jitter Measurements*

As part of the White Rabbit calibration [16], the offset of pulse per second (PPS) and clock signals are measured on an oscilloscope. The calibration tunes the internal signal delays of clock master and slave Ethernet I/O and can be used to minimize the PPS pulse offset between clock master and slave. For our purposes, the *variation* of the offset (jitter) is the more important property, since it affects the ADC clocking. A fixed offset is equivalent to cable delay from detector to digitizer and can be taken into account in offline processing of acquired data. With a White Rabbit Switch as the clock master and both a WR-LEN and a Pixie-Net XL as clock slaves, a fast oscilloscope reported jitter of 10-15 ps for the WR-LEN and ~100 ps for the Pixie-Net XL (standard deviation of 1002 offset measurements), both for the PPS signal and for a 10 MHz reference clock. Measuring jitter between the master PPS and the local clock edges, we find that the White Rabbit DAC controlled oscillator already has ~100 ps jitter, and the clock PLL for the ADCs increases it to ~300 ps. A second revision of the clock circuitry did not (yet) result in any improvements. However, the recovered Ethernet RX clock (which the White Rabbit tries to match with its clock adjustment scheme) is available from the FPGA with low jitter (17 ps) and can be connected to the ADC on DB01 in a workaround. The recovered clock does not adjust for phase delays as the White Rabbit clock would do, and is therefore not ideal for long cables that may exhibit varying transmission delays over time or temperature.

### C. *Time-of-Flight Style Measurements*

In time-of-flight style measurements (Fig. 3), the signal from a pulse generator (Agilent 33220A) is split between two

Pixie-Net XL. This mimics a setup of two detectors exposed to coincident radiation (e.g. from a $^{22}$Na source). The two Pixie-Net XL devices are synchronized through a network connection to a White Rabbit Switch acting as clock master. Both Pixie-Net XL units were instrumented with either the DB01 daughtercard digitizing at 75 MSPS or 125 MSPS or the DB02 daughtercard digitizing at 250 MSPS. Each Pixie-Net XL captured pulse waveforms with local triggers and recorded waveforms, 32 bit words of White Rabbit date/time (in 16 ns granularity), and processing clock time stamps (in 13.333 ns or 8 ns granularity). The time stamps were used for matching the event records, and an offline constant fraction algorithm was applied to the waveforms to determine the 50% level of the rising edge of the pulse to sub-sample precision. The result was a distribution of measured time differences, which were histogrammed and then peak fitted for the FWHM timing resolution shown in Fig. 4. While the time resolution for 2 channels within the same device was ~90 ps, resolutions were 550-750 ps for two devices in which ADCs where clocked with the White Rabbit output clock. However, when using the recovered Ethernet RX clock, the timing resolution with DB02 was 86 ps after some optimization of pulser amplitude and rise time.

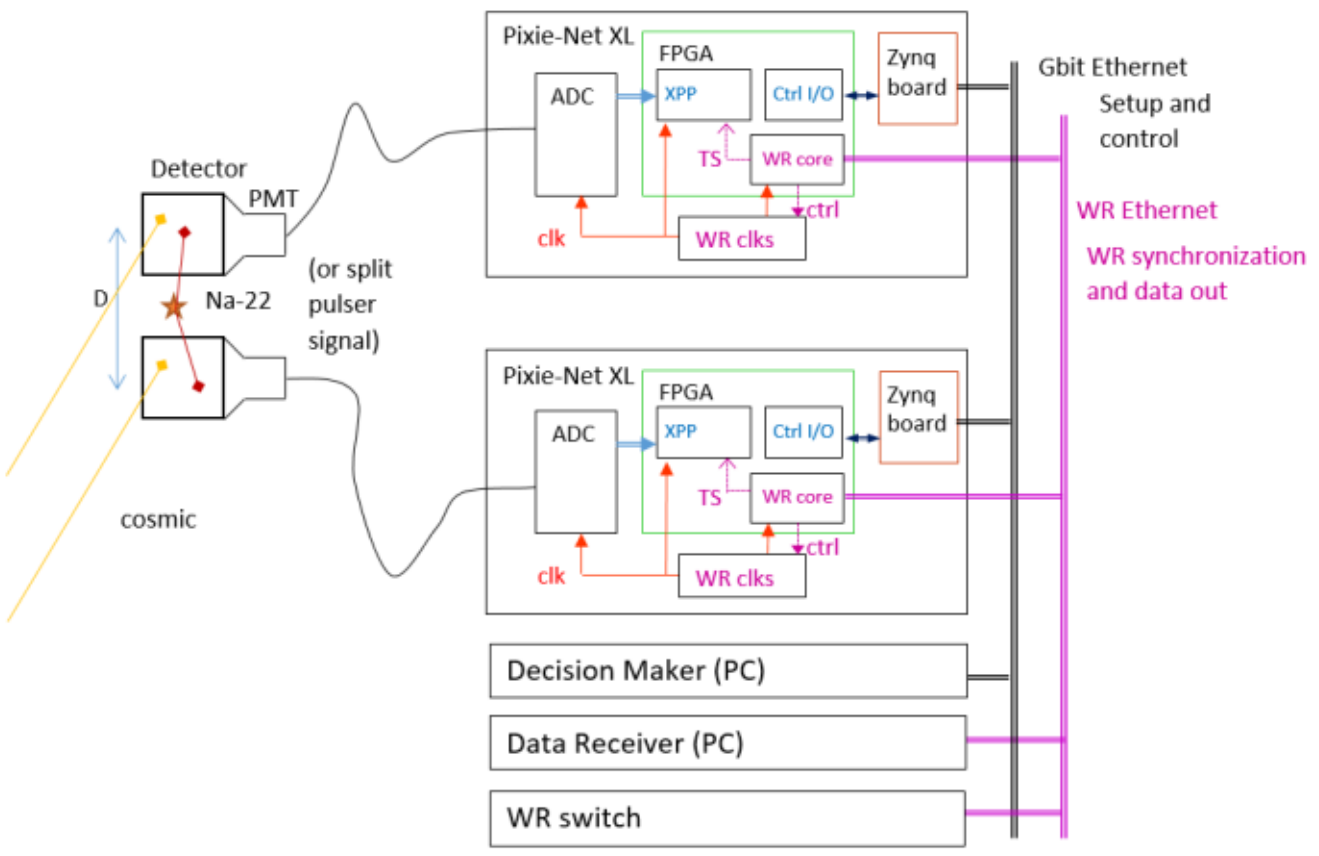

Fig. 3. Setup of time-of-flight and cosmic coincidence measurements. Detectors are separated by a distance *D*. Also shown are optional PCs for software triggering or data storage and the White Rabbit switch acting as a data router and clock master. Detectors can be replaced with a puser signal split to both ADCs. Separation of the setup and White Rabbit networks is for convenience, not a requirement.

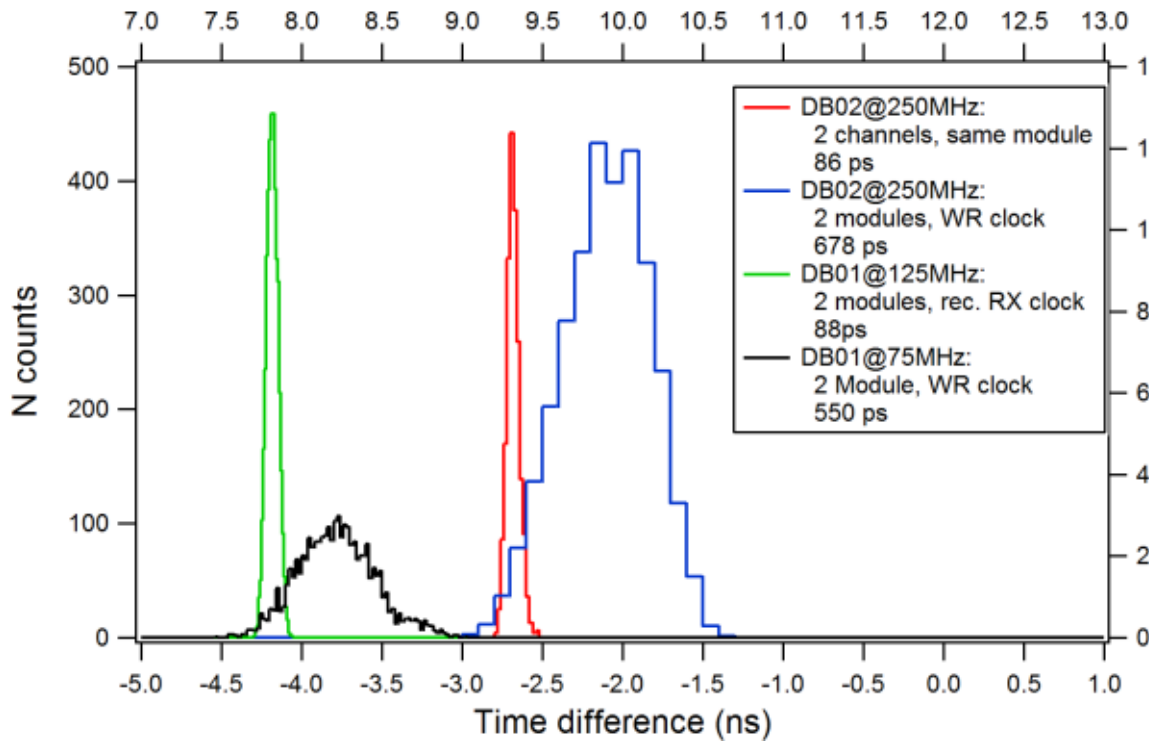

Fig. 4. Histogrammed measured time difference between two copies of pulses digitized in two Pixie-Net XL with DB01 and DB02.

## D. *Long Distance Detector Measurements with Cosmic Showers*

One of the attractive features of network time synchronization is that it functions over long distances using off-the-shelf network connections, rather than dedicated cabling that has to deal with cable losses for distances longer than a few meters. For example, the White Rabbit performance has been characterized with 5 km or more fiber lengths [17]. To test our system's timing performance over longer distances with coincident radiation, we would need extremely strong coincident sources or a particle accelerator. Fortunately, the latter are available free of charge in outer space and provide us with coincident radiation in the form of cosmic air showers that can spread over areas of $10^4$ m$^2$ or more for primary particles of $10^{15}$ eV or more [18]. The time of arrival difference between particles in the shower is set by the travel time from their creation point in upper atmosphere, and is therefore not expected to be in the picosecond range. Furthermore, for higher efficiency we use large NaI detectors (5 inch cubes) that are not designed for timing measurements. We here do not try to reproduce performance of large scale White Rabbit experiments like LHAASO [19]; rather this experiment is a good lab-scale test of the system in terms of long distance and long term operation.

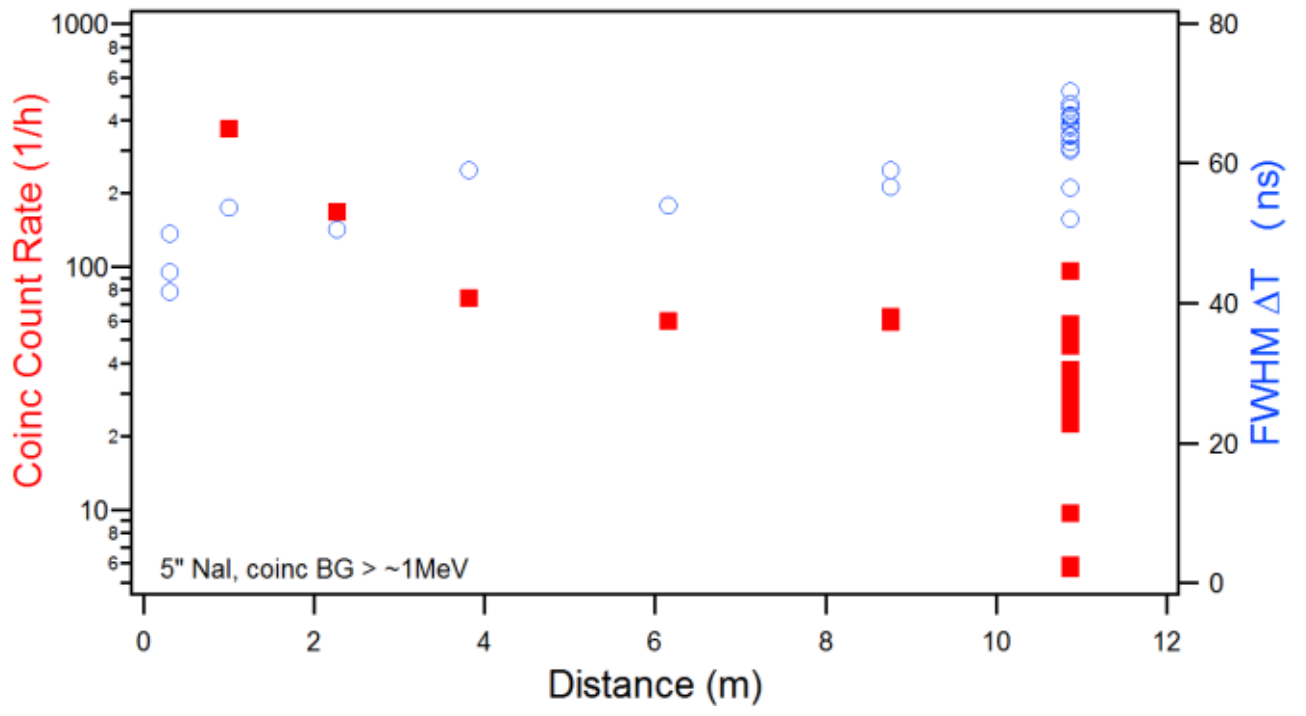

Fig. 5. Cosmic background coincidence rate as a function of detector distance, measured with detector electronics synchronized via White Rabbit.

The two NaI detectors were instrumented with two Pixie-Net XL units, one configured as White Rabbit master and one as slave. Detector distance *D* was increased from 30 cm to almost 11 m in steps of 1-2 m. Both Pixie-Net XL units were read out through the diagnostic data interface. In data acquisitions of typically 24 h, we recorded spectra of all detected energies and time stamped list mode data for only pulses with energies over about 1 MeV. At background rates of ~500 counts/s (all energies), this resulted in about 4.5 *million* records (>1 MeV) in 24 h and about 500 MB of data in each Pixie-Net XL. Matching events by White Rabbit time stamp and computing their time difference, we find *hundreds* of coincidences in 24h, illustrating the potential benefits of the software triggering scheme to make accept conditions in quasi real time and so reduce "waste data".

As expected from the earliest cosmic ray shower measurements [20], [21], the measured coincidence rate decreases with distance between the detectors (Fig. 5). While statistical errors are only a few percent, repeated measurements at the same distance vary significantly. Our measurements were performed at sea level with a 1000 ns coincidence windows and a detector area of approximately 160 $cm^2$, which makes comparison with historic data difficult. In particular, measurements in [20] were performed in the higher altitudes of the Swiss Alps, and unfortunately we were not able to travel there or to an equivalent location for measurements.

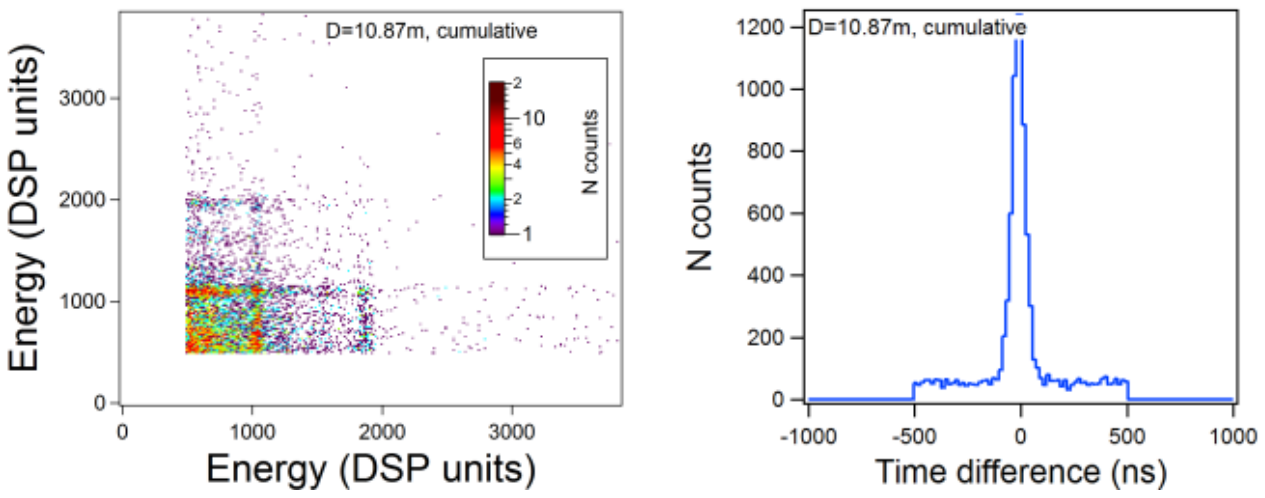

Fig. 6. Energy distribution (left) and time distribution (right) of coincidence events at 10.87m distance. The low E cutoff at 500 "DSP units" corresponds to ~ 1 MeV.

The time distribution of the coincidence cosmic particles formed a single peak with ~60 ns FWHM on a pedestal of random coincidences within the 1000 ns coincidence window (Fig. 6 right). As expected, this is not as good as in the pulser measurements due to the nature of the radiation and the slower detectors and PMTs. Plotting one energy against the other (Fig. 6 left) showed a structure of fixed-energy coincidence lines at roughly 2.2 and 3.8 MeV.

In later measurements, the software triggering was configured in two ways: 1) "accept all" where the DM is programmed to accept all trigger messages from either Pixie-Net XL, and each Pixie-Net XL is programmed to approve and store pulses matching time and device/channel number of the accept messages. 2) "coincidence only" where the DM is programmed to accept all trigger messages from either Pixie-Net XL (as before), and each Pixie-Net XL is programmed to approve and store pulses matching time *but not* device/channel number of the accept messages. This means locally detected pulses are only stored if the *other* device/channel detected a pulse at the same time. At a distance of ~0.5m, the total measured acceptance rate in "coincidence only" mode was then only 1-2% of the rate in "accept all" mode.

## E. *Data Output Rates*

The processing throughput was measured for 4 different modes of operation (Fig.7), using internal event and time counters in the firmware and a Berkeley Nucleonics random pulse generator. In list mode, event data consisting of energies, timestamps and 1 µs waveforms are output via the White Rabbit Ethernet or saved to SD card in diagnostic mode. In MCA-only mode, energies are only histogrammed in on-board (Zynq) memory. With all processing and decision making performed in the FPGA, the Pixie-Net can output ~300,000 events/s, roughly 80 Mbyte/s including UDP header words, before the internal buffering is overwhelmed and throughput drops drastically. If the Zynq is involved in the event acceptance, the exchange of metadata reduces the throughput to ~84,000 events/s. In diagnostic mode, using only the slow control bus between FPGA and Zynq, throughput drops to ~8,400 events/s. The maximum throughput strongly depends on the length of the captured waveform, and MCA-only processing with no data output can reach over 1 million events/s with very short energy filters (limited by the pulse generator).

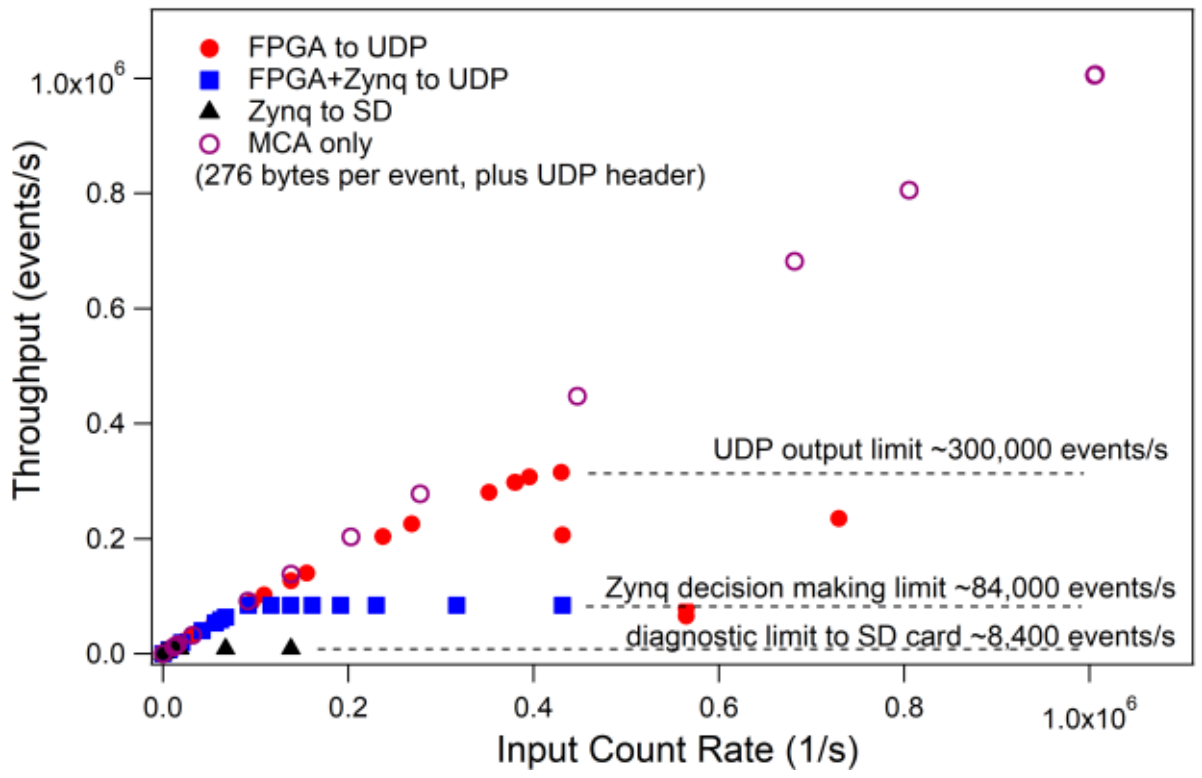

Fig. 7. Measured event data output rates for FPGA directed data flow to White Rabbit Ethernet (solid circles), Zynq directed accept/reject data flow to White Rabbit Ethernet (squares), and diagnostic mode to the Zynq SD card (triangles). Also shown is the pulse processing throughput in MCA-only mode (open circles). Each event consists of header with energy, timestamp, etc and a 1µs waveform, plus the UDP frame header.

Thus the data output flow can be managed either entirely inside the FPGA, or with acceptance tests applied by the Zynq processor, optionally with feedback from the DM. This allows users to operate at their preferred level of data reduction, from "store everything for later analysis" to "store only events with system-wide patterns of interest". The output data transmission is currently still in the form of UDP packets, which naturally raises questions of reliability due to the potential of lost packages. In the above tests, we only measured the outgoing data, with no regard to packets lost in transit or not properly received or stored on the UDP receiver, which would be a property of the network, not the Pixie-Net XL. (Other UDP drawbacks, such as the lack of handshaking and flow control, can be considered less relevant for pure data acquisition [22], and we specifically note that potential packet mis-ordering is not a problem here since each packet has a White Rabbit related time stamp as part of the event data.) A re-transmission scheme could relatively easily be added to the UDP packager in the FPGA, avoiding the complexities of a full TCP interface (and maintaining the ability to stream data to multiple receivers which TCP does not allow).

The hardware and the FPGA of the Pixie-Net XL are capable of 10G Ethernet rates. In an alternate firmware design, with the White Rabbit core replaced by a generic 10G core, the data output rate can thus in principle be increased by a factor of 10 at the cost of the White Rabbit synchronization. With preliminary firmware, we verified test data rates of at least 790 Mbyte/s generated and output by the FPGA, as reported by the traffic analysis of a 10G network switch.

## F. *Energy Resolution*

Since the Pixie-Net XL is designed as a detector readout module, we also measured its energy resolution with a high purity Germanium detector. At combined ~1000 counts/s (input rate), from multiple sources, we measured energy resolutions of ~0.14% at 1.3 MeV and just below 0.1% for 2.6 MeV (Fig. 8).

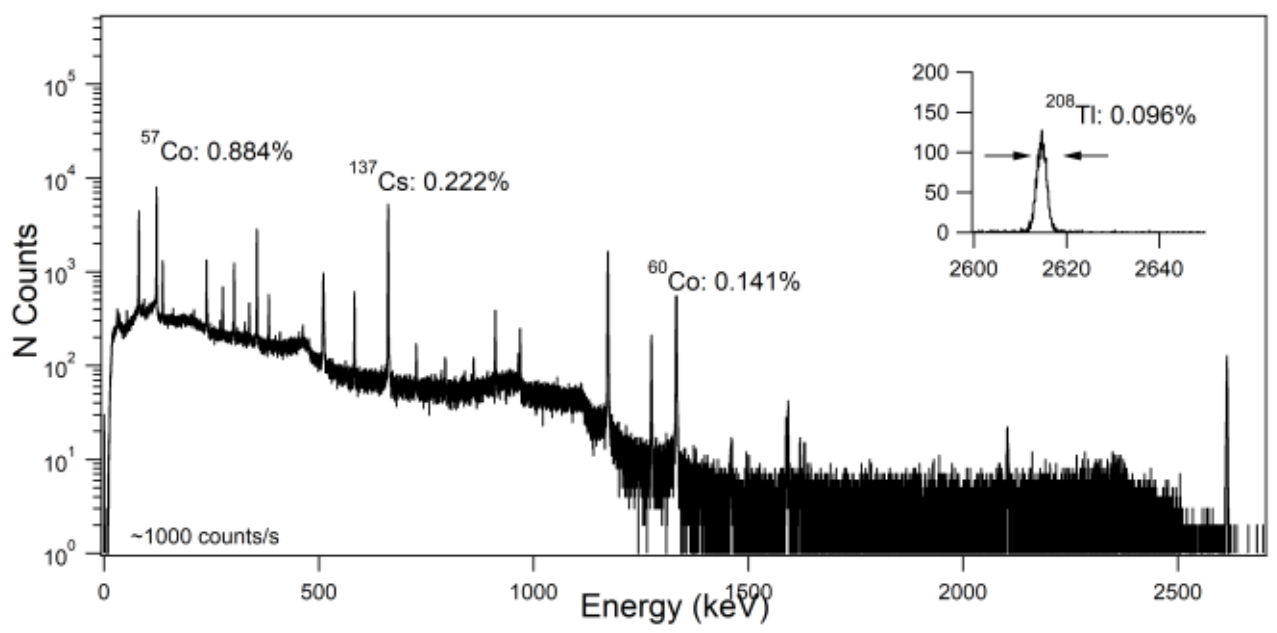

Fig. 8. Pixie-Net XL multi-source spectrum with HPGe detector (DB01).

## IV. Conclusion and Outlook

In summary, we implemented the White Rabbit time synchronization core in a new detector readout electronics module. Following the reference designs for hardware and firmware is quite straightforward, though special attention has to be given to the pinout of the FPGA's gigabit transceivers (ideally identical to the reference design to avoid reconstructing the transceiver firmware) and also the White Rabbit clocking circuitry for best jitter performance.

Time resolutions in preliminary measurements are 5-15 ps per software output and ~100ps per clock jitter measurements. In time-of-flight equivalent measurements with two modules, we obtained resolutions of 550-750 ps with the White Rabbit clock and 86 ps with the recovered RX Ethernet clock, compared to 90 ps for two channels in the same module using the same clock. Overall, timing resolutions are thus well below 1 ns. The method has been proven suitable for distances of more than 10 m, and there is no reason why it can not scale up with the same precision to distances set by the fiber optic network infrastructure, i.e. 5 km or more. The software triggering has been demonstrated to limit collection of unnecessary data. The White Rabbit Ethernet output data rate was measured to about 80 Mbyte/s, likely limited by the data packaging and buffering logic, not the White Rabbit core, and can be increased to 790 Mbytes/s or more with an upgrade to 10G Ethernet.

The test measurements were performed with preliminary versions of software, firmware, and hardware. In particular, there are sources of excessive jitter in the clock circuitry. The results are thus not yet quite as good as with commercially available devices or as reported in [17]. In future work, we will debug the jitter issues, finalize the firmware, and repeat the pulser measurements with fast radiation detectors.

## Acknowledgment

We thank the members of the White Rabbit OHWR support and discussion group for valuable assistance in implementing the White Rabbit firmware.

## References

[1] S. Akkoyun et al, "AGATA—Advanced GAmma Tracking Array", Nucl. Instrum. Meth. A, vol 668, pp. 26-58, 2012 doi.org/10.1016/j.nima.2011.11.081

[2] W. Hennig, H. Tan, M. Walby, P. Grudberg, A. Fallu-Labruyere, W. K. Warburton, C. Vaman, K. Starosta, D. Miller "Clock and Trigger Synchronization between Several Chassis of Digital Data Acquisition Modules", Nucl. Instrum. Meth. B, vol 261 pp. 1000–1004, 2007

[3] A. Kimura, M. Koizumi, Y. Toh, J. Goto and M. Oshima, "Performance of a data acquisition system for a large germanium detector array", International Conference on Nuclear Data for Science and Technology 2007, https://doi.org/10.1051/ndata:07400

[4] F. Alessio, S. Baron, M. Barros Marin, J.P. Cachemiche, F. Hachon, R. Jacobsson and K. Wyllie, "Clock and timing distribution in the LHCb upgraded detector and readout system" Journal of Instrumentation, Volume 10, February 2015, https://doi.org/10.1088/1748-0221/10/02/C02033

[5] Bing He, Ping Cao, De-Liang Zhang, Qi Wang, Ya-Xi Zhang, Xin-Cheng Qi, Qi An, "Clock distribution for BaF2 readout electronics at CSNS-WNS", 2017 Chinese Phys. C 41 016104, https://doi.org/10.1088/1674-1137/41/1/016104

[6] W. Hennig, V. Thomas, S. Hoover, O. Delaune, "Network Time Synchronization of the Readout Electronics for a New Radioactive Gas Detection System", IEEE Trans. Nucl. Sci., vol 66 no. 7, pp1182-1189 July 2019; https://doi.org/10.1109/TNS.2018.2885488

[7] S. Seifert, R Vinke, H. T. van Dam, H. Löhner, P. Dendooven, F. J. Beekman, D. R. Schaart., "Ultra precise timing with SiPM-based TOF PET scintillation detectors," in Proc. IEEE NSS/MIC Conf., Oct./ Nov. 2009, pp. 2329–2333. https://doi.org/10.1109/NSSMIC.2009.5402260

[8] M. Moszynski, M. Gierlik, M. Kapusta, A. Nassalski, T. Szczesniak, C. Fontaine, P. Lavoute, "New Photonis XP20D0 photomultiplier for fast timing in nuclear medicine", NIM A, 567 (2006), p31, https://doi.org/10.1016/j.nima.2006.05.054

[9] W. K. Warburton, W. Hennig, "New Algorithms For Improved Digital Pulse Arrival Timing With Sub-GSps ADCs", IEEE Trans. Nucl. Sci. Vol 64, No 12, pp. 2938-2950, Dec. 2017; DOI: 10.1109/TNS.2017.2766074

[10] *IEEE 1588-2008 - IEEE Standard for a Precision Clock Synchronization Protocol for Networked Measurement and Control Systems.* IEEE Standards Association [Online]. Available: https://standards.ieee.org/findstds/standard/1588-2008.html

[11] M. Lipiński, T. Włostowski, J. Serrano and P. Alvarez, "White Rabbit: a PTP application for robust sub-nanosecond synchronization," 2011 IEEE International Symposium on Precision Clock Synchronization for Measurement, Control and Communication, Munich, pp. 25-30, 2011. doi: 10.1109/ISPCS.2011.6070148

[12] R. Palit, S. Saha, J. Sethi, T. Trivedi, S. Sharma, B.S. Naidu, S. Jadhav, R. Donthi, P.B. Chavan, H. Tan, W. Hennig, "A high speed digital data acquisition system for the Indian National Gamma Array at Tata Institute of Fundamental Research" NIM A 680, 11 (2012), p90-96

[13] W. Hennig, S. Asztalos, D. Breus, K. Sabourov, W.K. Warburton, "Development of 500 MHz Multi-Channel Readout Electronics for Fast Radiation Detectors", IEEE Transactions on Nuclear Science, Vol. 57, No. 4, August 2010, p. 2365-2370

[14] Xillybus Lite for Zynq-7000: Easy FPGA registers with Linux. Xillybus Ltd. [Online]. Available: http://xillybus.com/xillybus-lite

[15] deepfifo: A drop-in standard FPGA FIFO with Gigabyte depth. Available: http://xillybus.com/tutorials/deep-virtual-fifo-download-source

[16] White Rabbit open hardware repository. [Online] Available: https://ohwr.org/project/white-rabbit

[17] M. Lipinski, T. Wlostowski, J. Serrano, P. Alvarez, "White Rabbit: a PTP application for robust sub-nanosecond synchronization", presented at the 2011 IEEE International Symposium on Precision Clock Synchronization for Measurement, Control and Communication, https://doi.org/10.1109/ISPCS.2011.6070148

[18] K.H. Kampert, A.A. Watson, "Extensive air showers and ultra high-energy cosmic rays: a historical review", EPJ H (2012) 37: 359. https://doi.org/10.1140/epjh/e2012-30013-x

[19] Q. Du, G. Gong, W. Pan, "A packet-based precise timing and synchronous DAQ network for the LHAASO project" NIM A 732 (2013) 448-492 http://dx.doi.org/10.1016/j.nima.2013.05.135

[20] P. Auger, P. Ehrenfest, R. Maze, J. Daudin, and Robley A. Fréon, "Extensive Cosmic-Ray Showers", Rev. Mod. Phys. 11, 288 – Published 1 July 1939. https://doi.org/10.1103/RevModPhys.11.288

[21] Schmeiser K. and Bothe, W. 1938. "Die harten Ultrastrahlschauer." Ann. Phys.,424:161.

[22] Kolhörster, W., Matthes, I., and Weber, E. 1938. "Gekoppelte Höhenstrahlen." Naturwissenschaften 26:576.

[23] Christensen, M.J., Richter, T, "Achieving reliable UDP transmission at 10 Gb/s using BSD socket for data acquisition systems", 2020 *JINST* **15** T09005, https://doi.org/10.1088/1748-0221/15/09/T09005